# Development of a focused-X-ray luminescence tomography (FXLT) system

Michael C. Lun[1], Wei Zhang[1], Yue Zhao[1], Jeffrey Anker[2], Wenxiang Cong[3], Ge Wang[3], Changqing Li[1*]

[1]Department of Bioengineering, School of Engineering, University of California, Merced, 5200 North Lake Road, Merced, CA 95340

[2]Department of Chemistry, Department of Bioengineering, Center for Optical Materials Science and Engineering Technology (COMSET), and Institute of Environment Toxicology (Cu-ENTOX), Clemson University, Clemson, SC, 29634

[3]Department of Biomedical Engineering, Biomedical Imaging Center, Center for Biotechnology and Interdisciplinary Studies, Rensselaer Polytechnic Institute, Troy, NY 12180

*Corresponding author: cli32@ucmerced.edu

**Abstract**

Biophotonics is an active research area in molecular imaging, genetic diagnosis and prognosis, with direct applicability in precision medicine. However, long-standing challenges of biophotonics are well known due to low signal-to-noise ratio and poor image quality, mainly due to strong optical scattering especially in deep tissues. Recently, X-ray luminescence computed tomography (XLCT) has emerged as a hybrid molecular imaging modality and shown great promises in overcoming the optical scattering in deep tissues. However, its high spatial resolution capacity has not been fully implemented yet. In this paper, with a superfine focused X-ray beam we design a focused-X-ray luminescence tomography (FXLT) system for spatial resolution better than 150 µm and molecular sensitivity of 2 µM. First, we describe our system design. Then, we demonstrate that the molecular sensitivity of FXLT is about 5 µM considering the emitted visible photons from background. Furthermore, we analyze the measurement time per scan from measured photons numbers with a fiber bundle-PMT setup, report numerical and experimental results. Finally, we specify the imaging system performance based on numerical simulations and physical measurements.

# 1. INTRODUCTION

## 1.1 Biomedical molecular imaging modalities

Biomedical molecular imaging provides functional details inside either humans or small animal bodies at the molecular and cellular level. It plays a critical role in disease diagnostics, therapy planning and monitoring, drug delivery, etc. Despite the significant improvements made in diverse biomedical molecular imaging modalities, each poses different limitations. Positron emission tomography (PET) is a suitable molecular imaging tool for both clinical and preclinical studies. Nevertheless, its best-reported spatial resolution is about 0.7 mm that is close to its intrinsic physical limits (1). Single photon emission computed tomography (SPECT) is another good molecular imaging tool but its measurement sensitivity is degraded by the pinholes (2). Fluorescence molecular tomography (FMT) and bioluminescence optical tomography (BOT) are widely used in molecular imaging because of their high sensitivity in detecting optical contrast agents, but their spatial resolution for deep targets is limited due to strong optical scattering (3, 4). Magnetic resonance imaging (MRI) has high spatial resolution but its measurement sensitivity is limited due to the negative contrast; additionally, the MRI scanner is very expensive (5). Computed tomography (CT) also provides high spatial resolution and thus it is often used to obtain anatomical structural information, however, its soft tissue contrast sensitivity is low (6). Tumors are often more than 10 mm deep under the skin. To the best of our knowledge, for tissues deeper than 10 mm, none of the aforementioned modalities can reconstruct the molecular biodistribution with a spatial resolution better than 200 micrometers and with high detection sensitivity. In this study, we design a focused-X-ray luminescence tomography (FXLT) imaging system that has such potentials to fill this gap and to provide the necessary and important high-resolution and high-sensitivity technology that is currently lacking.

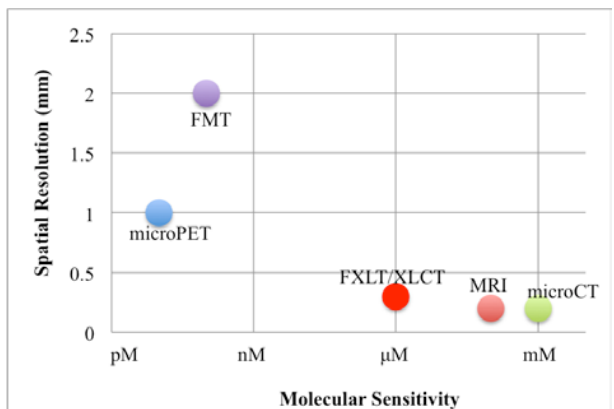

**Fig. 1** Approximate comparison of spatial resolution (y-axis) and molecular sensitivity (x-axis) of deep targets for different imaging modalities.

To have an approximate comparison among different molecular imaging modalities, we plot their performance in Fig. 1 based on reference (7) for FMT, PET and MRI, and reference (8) for microCT and X-ray luminescence computed tomography (XLCT). From Fig. 1, we see that the designed FXLT imaging system will provide a molecular imaging system with a spatial resolution better than 0.2 mm while having about two orders of magnitude better molecular sensitivity than MRI. If successful, the FXLT imaging system will provide a new and powerful platform for the nanomedicine and molecular imaging community.

## 1.2 XLCT background and history

Several groups, including our own, have explored the concept of XLCT imaging (9-13). Using a collimator to generate a fine X-ray pencil beam, XLCT could reconstruct two targets with an edge-to-edge distance of 0.4 mm in 5 mm deep optical scattering media (14). Without an X-ray collimator, conical X-ray beam based XLCT achieved a location accuracy about 1.5 mm (9). For "superficial" targets, XLCT can reach a spatial resolution of about 0.1 mm as demonstrated through numerical simulations (13). XLCT has also been proven to recover the nanophosphor concentration linearly (15). In addition, Zhang et al. have used multiple pinholes to speed up the XLCT imaging time and have also found that the radiation dose of XLCT is in a typical range of a CT scan (16). Recently, Lun et al. have demonstrated that XLCT was able to reconstruct a 21 mm deep target with a phosphor concentration of 0.01 mg/mL or 27 µM with a collimated X-ray beam scan scheme (8). The molecular sensitivity can be improved dramatically if we applied a focused X-ray beam based XLCT because the focused X-ray beam has three orders of magnitude higher photon density than a collimated X-ray beam (17). Furthermore, Zhang et al. have also proven that the spatial resolution of XLCT is about double the size of the X-ray beam diameter (18). Cong et al. have proposed the concept to use a dual-cone geometry of a focused X-ray beam to achieve high spatial resolution XLCT imaging and validated their idea with numerical simulations (19). However, until now, there is no reported XLCT imaging systems able to explore high spatial resolution imaging of up to 150 micrometers for X-ray excitable nanophosphors in deep tissues. In this study, we plan to build an XLCT imaging system, in which we use a superfine focused X-ray beam to achieve a spatial resolution of 150 micrometers for deep targets.

There are two types of XLCT imaging systems in terms of the excitation beam geometries. One is the conical X-ray beam based XLCT imaging (9, 20), in which a conical X-ray beam excites the whole mouse body. This approach can image the mouse in a very short time but the spatial resolution is compromised because there is no structural guidance from the X-ray beam in the XLCT reconstruction. Another is the pencil beam based XLCT imaging system (10, 13), in which a pencil beam X-ray is used to scan the object sequentially. The pencil beam approach generally requires a long measurement time (typically up to 1 hour) but can achieve sub-millimeter spatial resolution by using the pencil beam size and position as structural guidance in the XLCT image reconstruction. To obtain the high spatial resolution imaging for deep targets, we have focused on the superfine X-ray beam based XLCT imaging (18).

## 1.3 Nanophosphor development and applications.

Gadolinium oxysulfide (GOS) doped with either europium (Eu) or terbium (Tb) has a high cross-section for diagnostic energy X-rays and an excellent light yield, therefore it is often the phosphor of choice for X-ray detectors (21). Nanoscale X-ray excitable particles of GOS:$Eu^{3+}$ and other lanthanide-doped compounds have been successfully synthesized (22-26). The lanthanide cation in these compounds emits light in the red region due to $^5D_0 \rightarrow {}^7F_j$ radiative transitions, where j=4 corresponds to emission at wavelengths around 710 nm (26), which provides excellent tissue penetration for *in vivo* applications. The X-ray excitable nanoparticles can be coated with an amphiphilic polymer of octylamine-modified poly(acrylic acid) (27) or a plasmonic gold shell to improve the biocompatibility, then functionalized to target antibody-based proteins such as prostate-specific antigen (28). A plasmonic gold shell has been successfully added on a quantum dot core, still preserving its efficiency (29). Similarly a plasmonic gold shell could be coated on X-ray excitable nanophosphors (30), allowing widely researched gold labeling methods to be used with XLCT imaging (31). We and other groups have synthesized emission efficient Eu$^{3+}$ doped X-ray excitable nanophosphors (30, 32). Recently, Anker group at Clemson University has used X-ray excitable nanoparticles to monitor drug release by measuring the emission peak ratio (33-35). In addition, Chen group at the University of Texas Arlington have reported theranostic nanoparticles generating optical photons for imaging and singlet oxygen for therapy (36, 37).

### 1.4 Biological significance of the proposed XLCT imaging system

*Theranostic nanoparticle monitoring:* Prof. Wei Chen at the University of Texas Arlington, has synthesized copper-cysteamine (Cu-Cy) nanoparticles which can generate singlet oxygen directly when excited by X-ray photons; thus they are good for deep cancer treatment (36). Also, the Cu-Cy nanoparticles can emit bright optical photons with a wavelength peak at 650 nm (36). The Cu-Cy nanoparticle is an ideal molecular agent for the proposed FXLT imaging system to demonstrate the novel concept of theranostics, in which the FXLT system will image the fine biodistribution of Cu-Cy inside tumors for a precise X-ray dose delivery to make it possible for a tumor-specific therapeutic plan.

*Drug delivery monitoring:* There are two peak wavelengths in the nanophosphor emission spectrum when excited by high energy X-ray photons (30). Also, there is relatively low autofluorescence from tissues when excited by X-rays. These two unique features make the X-ray excitable nanophosphors a good candidate to monitor drug delivery quantitatively. Chen et al. have reported radioluminescent nanocapsules as the pH-triggered drug delivery vehicles and monitored the drug release with the emission ratios (35). The same group has also studied the multifunctional yolk-in-shell nanoparticles for pH-triggered drug release and reported the imaging with X-ray excitation (34). These nanoparticles have demonstrated the power of X-ray luminescence for drug delivery and monitoring. However, these studies were performed as two-dimensional (2D) imaging only. The designed FXLT imaging system will be an ideal platform to monitor the drug release inside tumors three-dimensionally and longitudinally.

*Cancer imaging:* Cancer vascularity has high permeability (38). Unfunctionalized nanoparticles will accumulate inside tumors through the leakage mechanism after intravenous injection. After adding a gold shell to increase their biocompatibility (30), the nanophosphors can be used to image tumors by the proposed FXLT system. Also, nanophosphors could also be labeled with epidermal growth factor receptor (EGFR) antibodies (32) to image the tumors' density of EGFR receptors at a superfine scale (39).

In this study, we will design the FXLT imaging system and demonstrate its feasibility from different perspectives including high spatial resolution with numerical simulations, molecular sensitivity with experimental measurements, and scan time with comparison study. This paper is organized as follows. Section 2 describes the designed FXLT imaging system, numerical simulation setup, and the molecular sensitivity measurement setup. The results are described in Section 3. Finally, we conclude the paper with discussions in Section 4.

## 2. METHODS

### 2.1 System Design

#### 2.1.1 Schematic of FXLT imaging system

Fig. 2 shows the CAD model of the proposed FXLT imaging system. We use an X-ray tube with a polycapillary lens (XOS, East Greenbush, NY) to focus the X-ray beam with a focal diameter of 55 micrometers. The X-ray tube with the polycapillary lens is mounted on a linear stage

(NLS8, Newmark Systems Inc.) that is fixed on a powerful rotary stage. For each rotary angle, the linear stage will move in total about 3 centimeters so that the focused X-ray beam will scan a single transverse section of the mice in less than 15 seconds per angular projection. The rotary frame is a heavy duty ring track from HepcoMotion (Devon, England), which has central bore diameter of 650 mm. A mouse to be imaged is placed on a transparent and thin stage that is mounted on a motorized linear stage. The focused X-ray beam scans the mouse linearly for specific angular projections. The angular projection number is flexible with a typical projection number of 6. At the opposite side of the mice, a single pixel X-ray detector is used to measure the transmitted X-ray photon intensity to monitor the focused X-ray. The single pixel X-ray sensor is fixed on a U-shape frame mounted on the X-ray tube's linear stage so that the X-ray detector moves together with the X-ray tube. An array of 8 fiber bundles is mounted circularly around the mouse and is fixed two millimeters away from the focused X-ray beam. The emitted luminescence photons on the mice body surface are collected by the 8 fiber bundles and delivered to an array of eight photomultiplier tubes (PMTs) as shown in Fig. 3. The PMTs work in a single photon counting mode. Two sets of optical bandpass filters following 8 PMTs are used to measure the emission photons at two different wavelengths simultaneously. At each wavelength, we have four measurement points for each linear scan step. The electrical signals from the PMTs will be amplified, filtered and then collected by a fast eight channel data acquisition (DAQ) board, which will be controlled by a computer.

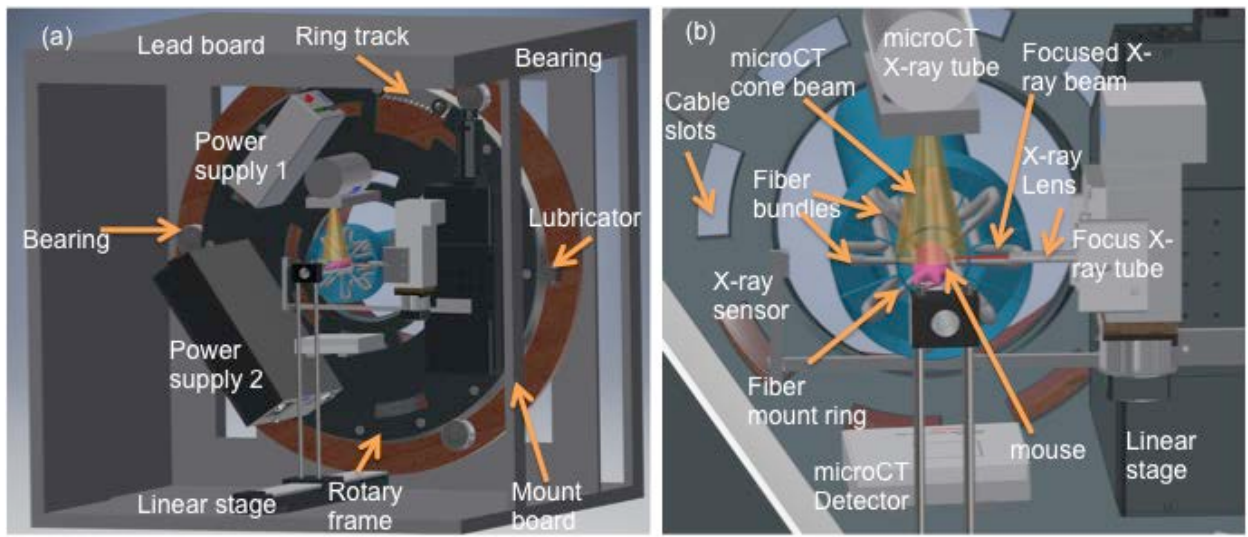

Fig. 2 (a) Right front view and (b) zoomed in view of the CAD model of the proposed FXLT imaging system.

As shown in Fig. 2, an X-ray tube (Jupiter 5000, Oxford Instruments, 50 kVp and 1 mA) and a flat panel X-ray detector (C7942CA-29, Hamamatsu) are mounted for microCT imaging of the imaged object. Before FXLT imaging, the linear stage places the object in the field of view (FOV) of the microCT that will take measurements at 360 angular projections with an angular step size of 1 degree. After the microCT scan, the linear stage moves the object to the FOV of the FXLT. A filtered back-projection with a Shepp-Logan filter is used to reconstruct the microCT images.

The FXLT imaging system, except the PMT arrays, amplifiers, filters and computer, is placed on an optics tables inside an X-ray shielding and light tight lead cabinet. The FXLT's X-ray tube is air-cooled. The heat from the X-ray tube will warm up the cabinet. At the same time, we design a heating system to blow warm air around the mice stage to keep its temperature around the normal mice body temperature. We also install a water cooling system with a heat radiator inside the cabinet. We design a temperature control system to warm up the stage if the temperature is low and to turn on the water cooling system if the temperature is too high.

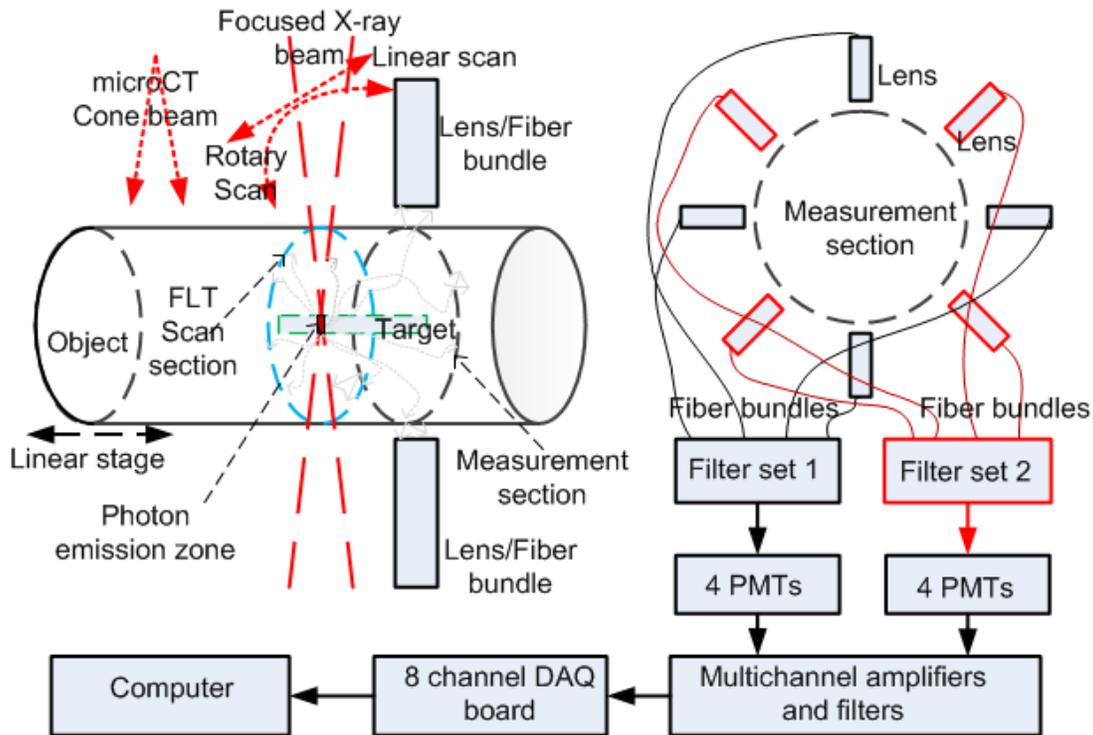

**Fig. 3** Scan scheme and 8 PMTs as optical photon detectors to measure photons at two different wavelengths simultaneously.

*2.1.2 Cone beam with a small beam angle*

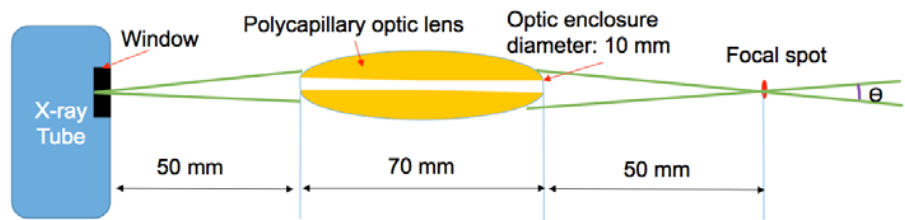

Fig. 4 Schematic of X-ray tube with a polycapillary optic lens. The convergence angle θ is less than 2 degrees.

In this study, we use a high power X-ray tube (Mo-anode) that is coupled with a polycapillary focusing lens optimized for Mo $K_\alpha$ radiation (X-Beam, XOS, East Greenbush, NY). The maximum power is 50 kV/50 W. The X-ray unit has a built-in cooling fan. Fig. 4 plots the approximate arrangement of the polycapillary optic lens. The distance from X-beam's window to its focus spot is 45 mm. The optic enclosure diameter is 10 mm. The output focal spot is less than 50 μm at Mo K$_α$ (17.4 keV). The beam size is less than 75 micrometers in a range up to 10 mm around the focal spot. The beam convergence angle is less than 2 degrees, which will be included in the reconstruction algorithm. The output intensity at 50 kV/50 W is larger than 2×10$^7$ photons/sec. The flux intensity is estimated to be 100 times larger than that of a collimator based X-ray beam.

*2.1.3 PMT arrays as photon detectors*

We use 4 photosensor modules with PMTs (H7422P-50, Hamamatsu) as the photon detectors to measure the emitted optical photons from the mice surface delivered by fiber bundles. Each module comes with a C-Mount adaptor, a heatsink, and a power supply. The efficient cooling reduces thermal noise from the photocathode and a high SNR is obtained even at extremely low light level. The H7422-50 is sensitive along a wide spectral range from 380 nm to 890 nm. Thus, the measurement time will be reduced dramatically compared the reported EMCCD camera based measurements. The output signals from the 4 photosensor modules are connected to a data acquisition (DAQ) board.

*2.1.4 Linear scan and angular projections of FXLT*

Our design allows us to have flexibility on the angular projection number. For the FXLT images, our simulation results have shown that measurements at 6 angular projections are sufficient to reconstruct complex XLCT images as described in the following sections. For each angular projection, we will have a continuous scan for one linear scan in 3 seconds. Thus, the total scan time for each transverse section depends on the angular projection number. For a typical angular projection number of 6, we need less than 1 minute for each transverse section scan if including the rotation time.

## 2.2 Molecular sensitivity of FXLT/XLCT imaging

Recently it has been reported that there are visible photons emitted when the X-ray beam excited water and air, even if the X-ray photon energy is less than the threshold energy for Cerenkov radiation (40). These photons will limit the molecular sensitivity of the proposed XLCT imaging system because these photons are from background of tissues and water. This limitation is the intrinsic limitation of the XLCT imaging. To estimate the intrinsic limitation, we have performed a comparison study of the emitted photons between the tissue-like phantom only and a tissue-like phantom with a GOS:Eu$^{3+}$ target (0.01 mg/mL). The phantom is composed of water, 1% agar, and 1% TiO$_2$ with a diameter of 25 mm. The focused X-ray beam excited the phantoms 5 mm below the top surface and through the phantom center. For the case with the target, the focused X-ray beam was through the target center too. During the measurement, the X-ray tube had a tube voltage of 50 kVp and a tube current of 1 mA. The emitted visible photons on the top surface were acquired by an EMCCD camera at its maximum gain with an exposure time of 5 seconds. The measurement system and the phantom were placed inside a light tight and X-ray shielding lead cabinet.

## 2.3 Measurement time of FXLT imaging

We have reported that measurements based a single fiber bundle with a PMT are sufficient for XLCT reconstruction when we used the focused X-ray beam (17). However, each linear scan step had a measurement time of 1 second. If we would like to reduce the measurement time, we need to measure how many emitted optical photons we can obtain from deep targets.

We performed a set of phantom experiments. For the phantom experiment, a background phantom was created using $TiO_2$ that measured 40 mm in height and 25 mm in diameter and was embedded with a single target that was 4.60 mm in diameter and contained 1.0 mg/mL of europium-doped gadolinium oxysulfide (GOS: $Eu^{3+}$) as shown below in Fig. 5. The optical properties of the phantom were an absorption coefficient ($\mu_a$) of 0.0072 $mm^{-1}$ and a reduced scattering coefficient ($\mu`_s$) of 1.00 $mm^{-1}$. The phantom was placed inside our XLCT imaging system and we collected the emitted optical photons when the focused X-ray beam passed through the center of the target at a scanning depth of 5 mm. The emitted optical photons from the phatom side surface was collected by a 2 meter long optical fiber bundle positioned 2 mm below the X-ray excitation section and was placed 90 degrees away from the target as seen in Fig. 5(b). The fiber bundle had an aperture diameter of 3 mm. The other end of the fiber bundle was positioned and fixed in front of a fan-cooled PMT (H7422P-50, Hamamatsu, Japan) that is driven by a high-voltage source (C8137-02, Hamamatsu) that measured the optical photons delivered by the fiber bundle. The electronic signal is then amplified by a broadband filter (SR445A, Stanford Research Systems, CA) using a total gain of 25. Afterwards, the signal is filtered using a low-pass filter (BLP-10.7+, cutoff frequency of 11 MHz, Mini-Circuits) and finally displayed using a high-speed 4 channel oscilloscope (MDO-3014, Tektronix, OR). During the X-ray excitation, the X-ray tube (XOS X-Beam Powerflux (Mo-anode), XOS, NY) was operated at the maximum power of 50 kVp and 1.0 mA (50 W) and had a focused X-ray beam with a focus spot of 98 µm.

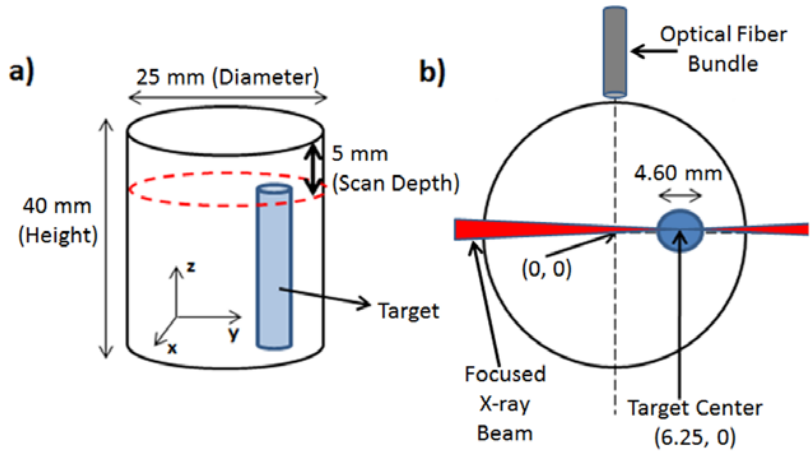

**Fig. 5** Phantom geometry for optical signal measurements using PMT. (a) Overall phantom geometry and (b) the top surface geometry.

*2.4 FXLT reconstruction models*

For each scanned section, we will calculate the sensitivity matrix at first from the known measurement locations. Then, we will interpolate the finite element mesh with a dense grid. The pixel size of the grid will be 25 µm × 25 µm. For each linear scan, we will find the superfine X-ray beam regions. Lastly, the system matrix will be calculated. The detailed steps are described in reference (41).

It is worth noting that the spatial resolution of the FXLT imaging system does not depend on the tissue optical properties. The optical properties will affect the imaged quantity. We can find the absorption coefficients and the reduced scattering coefficients of different tissues in mice from reference (42). The X-ray attenuation coefficient of tissues can be found in (12).

The proposed FXLT is a type of tomographic imaging, in which the reconstruction algorithms play a critical role. Especially, for sparse targets, the sparsity enhancement reconstruction algorithms such as $L^1$ and $L^q$ regularizations could reconstruct the images with reduced number of measurements that lead to less measurement time and a lower level of X-ray dose. The experimental data will be reconstructed with different regularizations and preconditioners. The reconstructed results will be compared. The best algorithms will be selected to further tune the experimental system design.

We have described the XLCT reconstruction algorithm using the X-ray pencil beam priors as in (41). Following similar modeling, we will develop the algorithms for FXLT as:

$$\hat{\mathbf{x}} = \min_{\mathbf{x} \geq 0} \Psi(\mathbf{x}) := \frac{1}{2} \| \mathbf{A}\mathbf{x} - \mathbf{b} \|^2 + \lambda \cdot \mathbf{R}(\mathbf{x}) \tag{1}$$

where $\mathbf{A}$ is the system matrix, $\mathbf{b}$ is the measurements, and $\boldsymbol{\lambda} \cdot \mathbf{R}(\mathbf{x})$ is the regularization term. The X-ray beam position and size information are included in the system matrix $\mathbf{A}$ as described in reference (41). We propose to compare the effects of different regularization methods as well as the effects of different preconditioning methods so that the best images can be reconstructed. We will solve Eq. 1 under a uniform optimization transfer framework. We follow the separable quadratic surrogate routine (43) to construct the surrogate function for the data-misfitting term. Regularizations are typically employed to counteract the ill-conditioning of $\mathbf{A}$ and suppress noises. We propose to compare the effects of different regularization methods as described in reference (44).

## 2.5 FXLT numerical simulations

To validate the feasibility of our designed FXLT imaging system, we have performed two sets of numerical simulations. For both simulation studies, we generated a cylindrical background phantom of 12 mm in diameter and 20 mm in height. The background phantom had an absorption coefficient ($\mu_a$) of 0.0072 mm$^1$ and a reduced scattering coefficient ($\mu_s$') of 0.72 mm$^{-1}$ to mimic the optical properties of mice tissue. Inside the background phantom we placed six cylindrical targets of 1.0 mg/mL GOS:Eu$^{3+}$ concentration as shown in Fig. 6. The 15 mm long targets each had diameters of 75 µm and were placed such that each target has an edge-to-edge distance of 75 µm. During the simulated scan, the emitted luminescent photons were collected by 4 detectors (fiber-PMT set-ups) that were positioned 2 mm below the scanned transverse section. We only performed scanning and reconstruction of a single transverse section of the phantom and targets which were discretized with a 2D grid having a pixel size of 10 µm × 10 µm.

In the first numerical simulation, we simulated the scanning scheme of the proposed FXLT imaging system and included the dual-cone geometry of the focused X-ray beam as well as X-ray attenuation and scattering (of 17.5 keV X-ray photons) in the forward model and reconstruction algorithm. For all simulation cases, we used a single dual-cone X-ray beam to scan the phantom at the depth of 5 mm from the phantom top surface and collected the emitted optical photons using 4 fiber-PMT set-ups surrounding the phantom 2 mm below the scanned section (each spaced 90º apart). In first numerical simulation, we incorporated as closely as we can, the actual dual-cone X-ray beam geometry provided by the manufacture (XOS, East Greenbush, NY) for our designed FXLT system as given below in Table 1. However, due to the constraints of our 2D grid, we round the focal spot to 50 µm in diameter and only incorporate the beam changes at ±2 mm (60 µm) as well as ±4 mm (70 µm). We placed the X-ray beam focal spot such that it would scan the center of our phantom during the FXLT scan. For this simulation study, we performed FXLT scans of different angular projections (3, 6, and 9) with angular step sizes of 60, 30, and 20 degrees respectively to determine the effect of different projection numbers on the spatial resolution. In addition, a second set of numerical simulations were performed to determine the effects of the X-ray beam size on the spatial resolution. In this case, we doubled the X-ray beam diameters used in the first simulation.

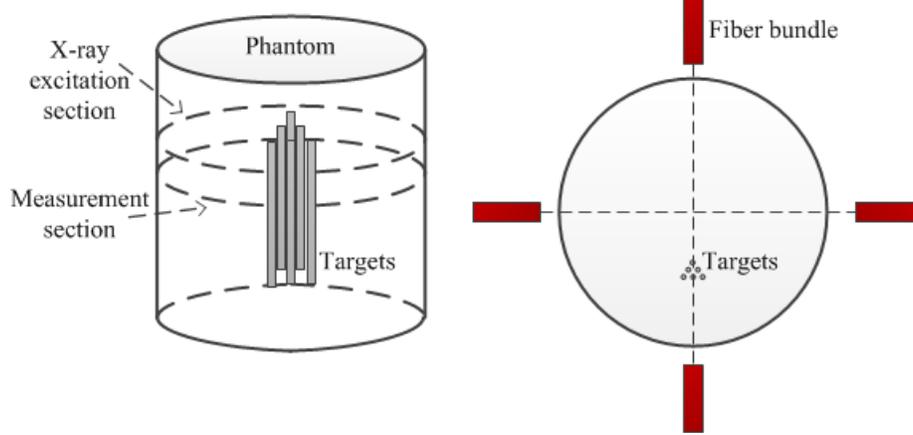

**Fig. 6** Phantom geometry and target positions for numerical simulations. Fiber bundles indicate the measurement locations.

**Table 1** Manufacture provided table of the output X-ray beam size.

| Distance from the focus | At Focus (25 mm OFD) | ±1mm | ±2mm | ±3mm | ±4mm | ±5mm | ±10mm |
|---|---|---|---|---|---|---|---|
| Output beam size at 17.5 keV (µm, FWHM) | 55 | 57 | 60 | 65 | 70 | 78 | 135 |

## 3. RESULTS

### 3.1 Molecular sensitivity of FXLT imaging

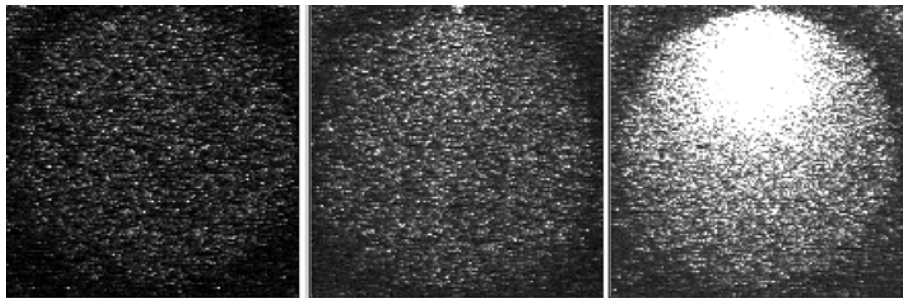

**Fig. 9** The emission photon pictures taken by the EMCCD camera for the cases of background (left, X-ray off), tissue-like phantom only (middle, X-ray on), and 0.01 mg/mL phosphor target in tissue-like phantom (right, X-ray on). All figures have the same grey scale with the brightness indicates more photons.

The raw EMCCD camera pictures are shown in Fig. 9. The leftmost figure is the background measurement picture when the X-ray tube is off. The middle figure is the measurement picture when the focused X-ray beam passed the phantom without target. The rightmost figure depicts the measurement picture when the focused X-ray beam passed the 0.01 mg/mL GOS:Eu$^{3+}$ target center. From the middle picture, we see that there are plenty of visible photons emitted from the background phantom. After we subtracted the background measurement (leftmost picture) from the measurements (middle and rightmost pictures), we have calculated that the photon intensity from the 0.01 mg/mL GOS:$Eu^{3+}$ target is 14 times brighter than the emission from the tissue-like phantom. The molar concentration of 0.01 mg/mL GOS:$Eu^{3+}$ is 27 µM. If the future synthesized nanophosphors have the similar emission efficiency, the molecular sensitivity of XLCT imaging in deep tissues will be about 27/14 ≈ 2 µM. A recent paper by Chen et. al. has demonstrated that nanoscale nanophosphors could have a luminescence intensity up to 40% of commercial microphosphors (45). The estimated molecular sensitivity of FXLT is about 2/0.4 = 5 µM.

### 3.3 Measurement time of FXLT imaging

For the data acquisition using the oscilloscope, we acquired data using a total measurement time of 2 milliseconds (ms) from which we acquired 10 million data points. The oscilloscope images for the phantom experiment using 1.0 mg/mL is plotted in Fig. 10. Fig. 10(a) shows the signal acquired when the X-ray tube is off and not exciting our target, from which we see that there are no photon peaks. Fig. 10(b) shows the signal we obtain when the X-ray tube is on and passing through the center of the target. We can see clearly a large signals obtained. To quantify the signals, we counted the number of peaks after subtracting a threshold value determined using the image in Fig. 10(a). After subtraction of the background signal, the total number of peaks was summed. When the X-ray beam is on and passing the center of the 1.0 mg/mL target, the total number of pulses with values exceeding the threshold increases to 432,930.

The emission photon number is proportional to the measurement time, phosphor concentration, target size, and X-ray photon flux. If we use a 75 µm X-ray beam to scan a 75 µm diameter target with a concentration of 0.01 mg/mL and a measurement time 0.05 second per 75 µm (or 1.5 mm/s), which is the worst scenario, we can estimate that the measured emission photon number is around 164. We need 15 seconds per angular projection scan and 2 minutes per section, including the rotating time between each angular projection, if six angular projection measurements are performed.

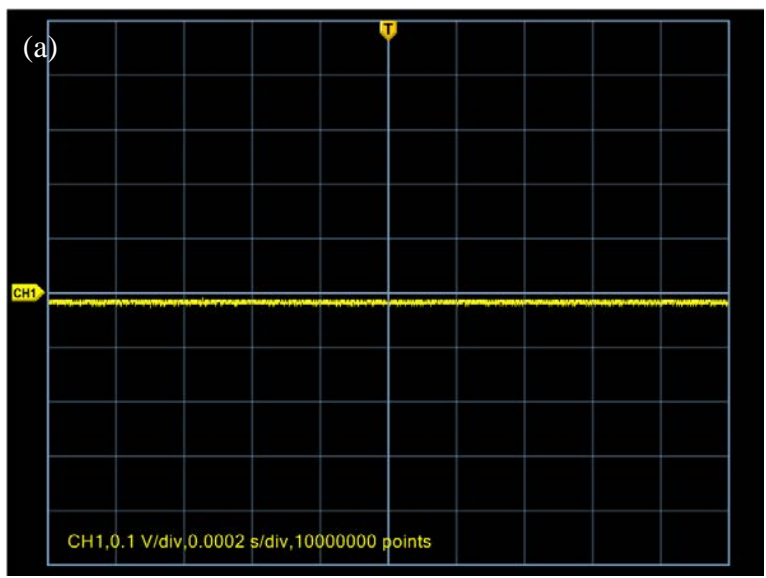

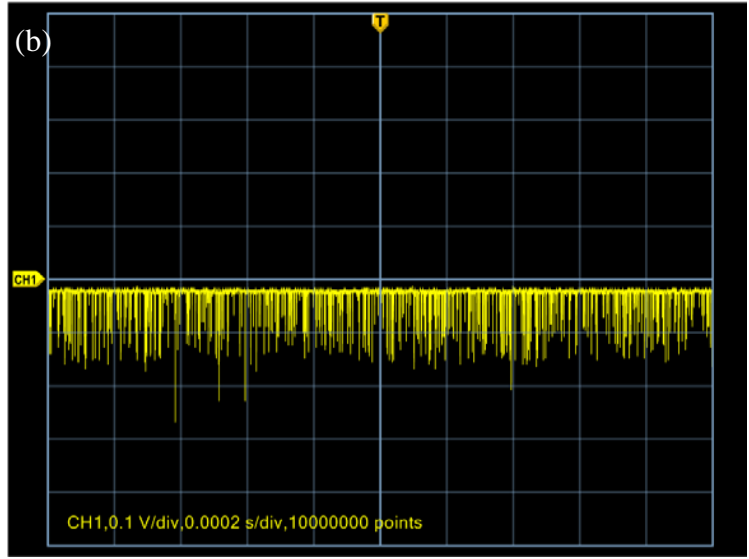

**Fig. 10** Oscilloscope images showing the signal obtained when (a) the X-ray tube is off and (b) when the X-ray tube is on.

## *3.4 Effect of projection number and beam size on spatial resolution*

For the image reconstruction, the system matrix generated by the forward model was interpolated to the fine 2D grid. During reconstruction, the $L^1$ regularization method was applied in the Majorization-Minimization reconstruction framework to reconstruct the images. Fig. 11 below shows the results of the numerical simulations of different angular projections of the proposed FXLT imaging system from which we can see the six targets have been successfully reconstructed and can be separated for all cases. The green circles in the figure represent the true target locations. We calculated the DICE similarity coefficients for each of the reconstructed cases and for each projection number of 3, 6, and 9 projections, the corresponding DICE was calculated using the full width ten percent maximum (FW10%M) to be 35.2490, 46.0967, and 49.3333 % respectively. In addition, the size of each of the two targets in the middle row was also calculated using the FW10%M and the results shown in Table 2. It should be noted that with our numerical simulation setup, it is impossible to obtain a perfect target size (0.075mm) since our grid is only 10 μm in size. Our results indicate that we can perform FXLT scans with as little as 3 angular projections and that we can improve the spatial resolution and imaging quality using more angular projections if required with the proposed imaging system.

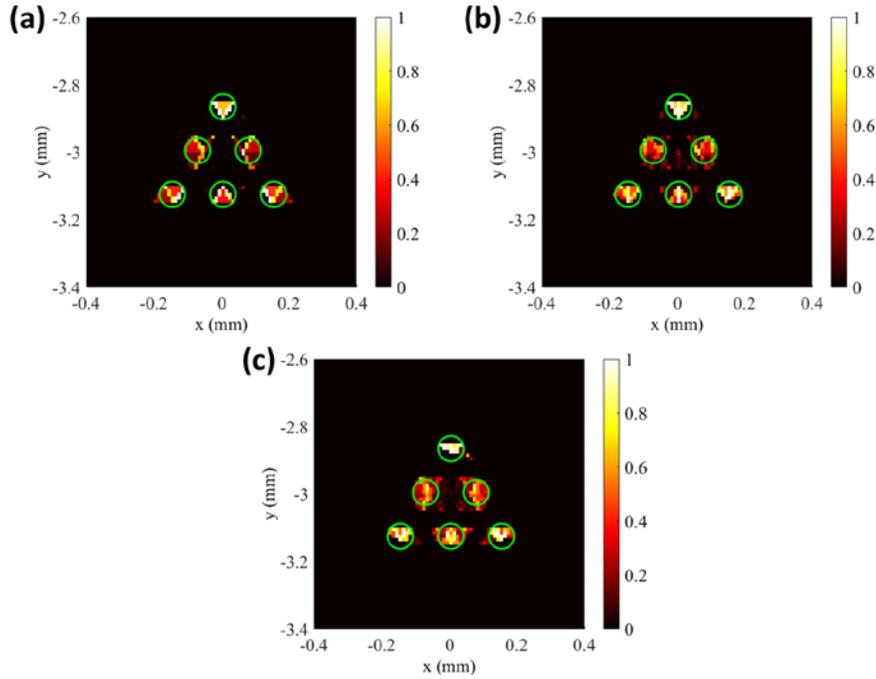

**Fig. 11** Results of numerical simulation determining the effects of projection number on spatial resolution. The green circle represents the true target regions. Reconstructed images at (a) 3 Projections, (b) 6 Projections, and (c) 9 Projections.

**Table 2** Calculated target sizes for different angular projections.

| # Angular Projections | Left Target Size (mm) | Right Target Size (mm) |
|---|---|---|
| 3 | 0.060 | 0.060 |
| 6 | 0.060 | 0.070 |
| 9 | 0.080 | 0.080 |

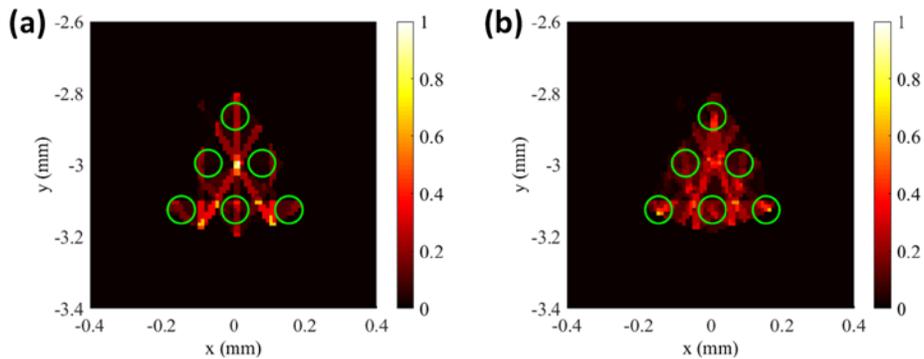

**Fig. 12** Results of numerical simulation of the effect of beam size. Reconstructed images using (a) 6 Projections and (b) 9 Projections. FXLT reconstruction failed when the beam size is large.

The second set of numerical simulations performed was to investigate the effects of the X-ray beam size on the imaging quality and spatial resolution of the proposed FXLT imaging system. This simulation was set-up exactly the same as the previous simulation, except that the

X-ray beam diameter used was twice the size. In this set-up we used an X-ray focal diameter of 100 μm and set the X-ray beam diameter to be 120 μm at ±2 mm away from the focal spot and 140 μm at ±4 mm away from focal spot. For this simulation, we performed scans using both 6 and 9 angular projections using angular step sizes of 30 and 20 degrees respectively. The results of the numerical simulation are shown above in Fig. 12. It can be seen that even at 9 angular projections, the targets cannot be resolved successfully. For the angular projections used (6 and 9), the corresponding DICE for the middle row targets were calculated to be 9.780 and 22.6138 % respectively. Compared to the earlier results in Figs. 11(b, c), the image quality is heavily degraded by using a larger X-ray beam and the targets cannot be separated. These results are consistent with our previous findings in reference (18) where we found that the spatial resolution of XLCT is about double the size of the X-ray beam diameter used.

### *3.5 FXLT imaging system performance estimation*

*Spatial resolution:* The numerical simulations in Section 3.4 indicates that the proposed FXLT imaging system is able to achieve a spatial resolution of 150 micrometers if we used the focused X-ray beam with a maximum diameter of 75 micrometers. The achievable spatial resolution is double size of the X-ray beam diameter, which is consistent with our previous report (18). It is worth noting that the spatial resolution is the spatial resolution in a transverse section. The spatial resolution along the axial direction depends on the X-ray beam size and the linear stage. Usually, a typical linear stage can have an accuracy of less than 10 micrometers. Thus, the spatial resolution will also depend on the X-ray beam size. We expect the spatial resolution along the axial direction is 75 micrometers.

*Molecular sensitivity:* As discussed in Section 3.2, the estimated minimum molecular sensitivity of the proposed FXLT imaging is about 5 µM when we image the nanophosphor deeply embedded (up to 10 mm) inside mice. It has been reported that the X-ray luminescence photons from background weigh more in blue wavelength range (46), where alpha particles were used as excitation source. They believe the similar emission spectra would be obtained for X-ray photon excitation of water. If we apply a bandpass filter to block most background photons, we might achieve lower molecular sensitivity.

*Scan time:* Our results in Section 3.3 indicate that we need at least 15 seconds for a linear scan to cover 22 mm, a typical size of the mouse trunk. From Section 3.4, we see that we can reconstruct good FXLT images with measurements at six angular projections with 4 fiber bundles as detectors. Thus, the minimum scan time will be 90 seconds per transverse section scan by FXLT. If we include the rotation time (around 5 seconds) between each angular projection, the total scan time per transverse section imaging will be around 120 seconds.

*X-ray dose:* From our previous paper (17), we see that the radiation dose per 12 minutes scan with the focused X-ray beam was 6.8 cGy. We are able to reduce the scan time to 90 seconds (exclude the rotary time). Thus we can estimate the radiation dose per transverse section FXLT imaging is about 6.8/(12/1.5)=0.85 cGy, which is in a range of typical CT imaging.

### **4. DISCUSSIONS AND CONCLUSION**

We have estimated the minimum or intrinsic molecular sensitivity of FXLT imaging by measuring the signal to noise ratio with a phantom. It would be more accurate if we used mice tissues to estimate the background luminescence signals. And, we have not considered the application of bandpass filters to block the background luminescence signals. If a bandpass filter was used to measure the emission photons from nanophosphors only, then we should have better molecular sensitivity than we estimated in this study.

The total photon number per linear scan step is the key factor in estimating the FXLT scan time. We have performed the photon number measurements with a tissue-mimic phantom composted of $TiO_2$ as scatters. $TiO_2$ has higher attenuation coefficient of X-ray photon than the Intralipid. If we used Intralipid as the optical scatterer, the measured photon number would be higher.

In sum, we have designed an FXLT imaging system with 150 µm spatial resolution and several µM molecular sensitivity for deeply embedded targets inside mice. We have estimated the FXLT imaging system performance with phantom experiments and numerical simulations. We have estimated that the scan time per transverse section imaging is about 120 seconds and the radiation dose will be around a typical CT scan range. Overall our results have indicated that the proposed FXLT imaging system has great potential to be a powerful tool for the molecular imaging research community.

**ACKNOWLEDGEMENT**

This work is supported in part by Grant (R03 EB022305) from the National Institute of Health (NIH).

**REFERENCES:**

1. Yang YF, Wu YB, Qi JY, James SS, Du HN, Dokhale PA, Shah KS, Farrell R, Cherry SR. A prototype PET scanner with DOI-encoding detectors. Journal of Nuclear Medicine. 2008;49(7):1132-40.
2. Peremans K, Cornelissen B, Van den Bossche B, Audenaert K, Van de Wiele C. A review of small animal imaging planar and pinhole spect Gamma camera imaging. Vet Radiol Ultrasoun. 2005;46(2):162-70.
3. Wang G, Cong WX, Durairaj K, Qian X, Shen H, Sinn P, Hoffman E, McLennan G, Henry M. In vivo mouse studies with bioluminescence tomography. Optics Express. 2006;14(17):7801-9.
4. Ntziachristos V, Ripoll J, Wang LHV, Weissleder R. Looking and listening to light: the evolution of whole-body photonic imaging. Nat Biotechnol. 2005;23(3):313-20.
5. Caravan P. Strategies for increasing the sensitivity of gadolinium based MRI contrast agents. Chem Soc Rev. 2006;35(6):512-23.
6. Beyer T, Townsend DW, Brun T, Kinahan PE, Charron M, Roddy R, Jerin J, Young J, Byars L, Nutt R. A combined PET/CT scanner for clinical oncology. Journal of Nuclear Medicine. 2000;41(8):1369-79.
7. Levin CS. New imaging technologies to enhance the molecular sensitivity of positron emission tomography. P Ieee. 2008;96(3):439-67.
8. Lun MC, Zhang W, Li CQ. Sensitivity study of x-ray luminescence computed tomography. Applied Optics. 2017;56(11):3010-9.
9. Liu X, Liao QM, Wang HK. In vivo x-ray luminescence tomographic imaging with single-view data. Optics Letters. 2013;38(22):4530-3.
10. Li C, Di K, Bec J, Cherry SR. X-ray luminescence optical tomography imaging: experimental studies. Optics Letters. 2013;38(13):2339-41.
11. Badea CT, Stanton IN, Johnston SM, Johnson GA, Therien MJ. Investigations on x-ray luminescence CT for small animal imaging. Proc Spie. 2012;8313.
12. Cong WX, Shen HO, Wang G. Spectrally resolving and scattering-compensated x-ray luminescence/fluorescence computed tomography. J Biomed Opt. 2011;16(6).

13. Pratx G, Carpenter CM, Sun C, Xing L. X-ray luminescence computed tomography via selective excitation: a feasibility study. IEEE Trans Med Imaging. 2010;29(12):1992-9.
14. Zhang W, zhu D, Zhang K, Li C. Microscopic x-ray luminescence computed tomography. SPIE BiOS. 2015;93160M:93160M-6.
15. Carpenter CM, Sun C, Pratx G, Rao R, Xing L. Hybrid x-ray/optical luminescence imaging: Characterization of experimental conditions. Med Phys. 2010;37(8):4011-8.
16. Zhang W, Zhu D, Lun M, Li CQ. Multiple pinhole collimator based X-ray luminescence computed tomography. Biomed Opt Express. 2016;7(7):2506-23.
17. Zhang W, Lun M, Li CQ. Fiber based fast sparse sampling x-ray luminescence computed tomography. Proc SPIE. 2017;10057:1005704.
18. Zhang W, Zhu D, Lun M, Li CQ. Collimated superfine x-ray beam based x-ray luminescence computed tomography. Journal of X-ray science and technology. 2017;25:(In Press).
19. Cong WX, Pan ZW, Filkins R, Srivastava A, Ishaque N, Stefanov P, Wang G. X-ray micromodulated luminescence tomography in dual-cone geometry. J Biomed Opt. 2014;19(7).
20. Zhang GL, Liu F, Liu J, Luo JW, Xie YQ, Bai J, Xing L. Cone Beam X-ray Luminescence Computed Tomography Based on Bayesian Method. Ieee T Med Imaging. 2017;36(1):225-35.
21. Gambaccini M, Taibi A, DelGuerra A, Marziani M, Tuffanelli A. MTF evaluation of a phosphor-coated CCD for x-ray imaging. Physics in Medicine and Biology. 1996;41(12):2799-806.
22. Xing MM, Cao WH, Pang T, Ling XQ, Chen N. Preparation and characterization of monodisperse spherical particles of X-ray nano-phosphors based on $Gd_2O_2S$:Tb. Chinese Sci Bull. 2009;54(17):2982-6.
23. Thirumalai J, Chandramohan R, Valanarasu S, Vijayan TA, Somasundaram RM, Mahalingam T, Srikumar SR. Shape-selective synthesis and opto-electronic properties of $Eu^{3+}$-doped gadolinium oxysulfide nanostructures. Journal of Materials Science. 2009;44(14):3889-99.
24. Tian Y, Cao WH, Luo XX, Fu Y. Preparation and luminescence property of $Gd_2O_2S$ : Tb X-ray nano-phosphors using the complex precipitation method. Journal of Alloys and Compounds. 2007;433(1-2):313-7.
25. Wang HY, Wang RJ, Sun XM, Yan RX, Li YD. Synthesis of red-luminescent $Eu^{3+}$-doped lanthanides compounds hollow spheres. Materials Research Bulletin. 2005;40(6):911-9.
26. Pires AM, Davolos MR, Stucchi EB. $Eu^{3+}$ as a spectroscopic probe in phosphors based on spherical fine particle gadolinium compounds. Int J Inorg Mater. 2001;3(7):785-90.
27. Wang S, Jarrett BR, Kauzlarich SM, Louie AY. Core/shell quantum dots with high relaxivity and photoluminescence for multimodality imaging. J Am Chem Soc. 2007;129(13):3848-56.
28. Lutz B, Dentinger C, Sun L, Nguyen L, Zhang JW, Chmura AJ, Allen A, Chan S, Knudsen B. Raman nanoparticle probes for antibody-based protein detection in tissues. J Histochem Cytochem. 2008;56(4):371-9.
29. Jin YD, Gao XH. Plasmonic fluorescent quantum dots. Nat Nanotechnol. 2009;4(9):571-6.
30. Sudheendra L, Das GK, Li C, Stark D, Cena J, Cherry S, Kennedy IM. $NaGdF_4$:$Eu^{3+}$ Nanoparticles for Enhanced X-ray Excited Optical Imaging. Chemistry of Materials. 2014;26(5):1881-8.
31. Hainfeld JF, Powell RD. New frontiers in gold labeling. J Histochem Cytochem. 2000;48(4):471-80.

32. Sun C, Pratx G, Carpenter CM, Liu HG, Cheng Z, Gambhir SS, Xing L. Synthesis and Radioluminescence of PEGylated Eu3+-doped Nanophosphors as Bioimaging Probes. Adv Mater. 2011;23(24):H195-H9.
33. Moore TL, Wang FL, Chen HY, Grimes SW, Anker JN, Alexis F. Polymer-Coated Radioluminescent Nanoparticles for Quantitative Imaging of Drug Delivery. Adv Funct Mater. 2014;24(37):5815-23.
34. Chen HY, Qi B, Moore T, Wang FL, Colvin DC, Sanjeewa LD, Gore JC, Hwu SJ, Mefford OT, Alexis F, Anker JN. Multifunctional Yolk-in-Shell Nanoparticles for pH-triggered Drug Release and Imaging. Small. 2014;10(16):3364-70.
35. Chen H, Moore T, Qi B, Colvin DC, Jelen EK, Hitchcock DA, He J, Mefford OT, Gore JC, Alexis F, Anker JN. Monitoring pH-Triggered Drug Release from Radioluminescent Nanocapsules with X-ray Excited Optical Luminescence. Acs Nano. 2013;7(2):1178-87.
36. Ma L, Zou XJ, Chen W. A New X-Ray Activated Nanoparticle Photosensitizer for Cancer Treatment. J Biomed Nanotechnol. 2014;10(8):1501-8.
37. Ma L, Chen W, Schatte G, Wang W, Joly AG, Huang YN, Sammynaiken R, Hossu M. A new Cu-cysteamine complex: structure and optical properties. J Mater Chem C. 2014;2(21):4239-46.
38. Karathanasis E, Chan L, Karumbaiah L, McNeeley K, D'Orsi CJ, Annapragada AV, Sechopoulos I, Bellamkonda RV. Tumor Vascular Permeability to a Nanoprobe Correlates to Tumor-Specific Expression Levels of Angiogenic Markers. Plos One. 2009;4(6).
39. Weis SM, Cheresh DA. Tumor angiogenesis: molecular pathways and therapeutic targets. Nat Med. 2011;17(11):1359-70.
40. Yamamoto S, Koyama S, Komori M, Toshito T. Luminescence imaging of water during irradiation of X-ray photons lower energy than Cerenkov- light threshold. Nucl Instrum Meth A. 2016;832:264-70.
41. Li C, Martinez-Davalos A, Cherry SR. Numerical simulation of x-ray luminescence optical tomography for small-animal imaging. J Biomed Opt. 2014;19(4):46002.
42. Alexandrakis G, Rannou FR, Chatziioannou AF. Effect of optical property estimation accuracy on tomographic bioluminescence imaging: simulation of a combined optical-PET (OPET) system. Physics in Medicine and Biology. 2006;51(8):2045-53.
43. Erdogan H, Fessler JA. Ordered subsets algorithms for transmission tomography. Physics in Medicine and Biology. 1999;44(11):2835-51.
44. Zhu DW, Li CQ. Nonuniform update for sparse target recovery in fluorescence molecular tomography accelerated by ordered subsets. Biomed Opt Express. 2014;5(12):4249-59.
45. Chen HY, Wang FL, Moore TL, Qi B, Sulejmanovic D, Hwu SJ, Mefford OT, Alexis F, Anker JN. Bright X-ray and up-conversion nanophosphors annealed using encapsulated sintering agents for bioimaging applications. J Mater Chem B. 2017;5(27):5412-24.
46. Yamamoto S, Komori M, Koyama S, Toshito T. Luminescence imaging of water during alpha particle irradiation. Nucl Instrum Meth A. 2016;819:6-13.